%\magnification=1200
%\hsize=16 true cm
%\tolerance 2200
%\vsize=23.5 true cm
%\nopagenumbers      %  pagenumbers
%\topskip=1 truecm
%\normalbaselineskip=12pt
%\normalbaselines
%\headline={\tenrm\hfil\folio\hfil}
%\raggedbottom
%\abovedisplayskip=3mm
%\belowdisplayskip=3mm
%\abovedisplayshortskip=1mm
%\belowdisplayshortskip=2mm
\def\ref#1{$[#1]$}
%\vglue 2.5 true cm
%\vglue 2.5 true cm
\vsize 21.6 truecm
\hsize 15.2 truecm

%\font\twelvebf=cmbx12
\font\twelvebf=cmbx12
\font\eightbf=cmbx8
\font\eightrm=cmr8

\def\newline{\hfill\break}

\centerline{\twelvebf  STATISTICAL MECHANICS AND CAPACITY-APPROACHING } 

\vglue 1 true cm

\centerline{\twelvebf   ERROR-CORRECTING CODES }

\vglue 1.5 true cm
\centerline{{\bf  Nicolas Sourlas}} 
\medskip
\centerline{Laboratoire de Physique Th\'eorique de l' Ecole Normale 
Sup\'erieure  \footnote {*}  
{UMR 8549, Unit{\'e}   Mixte de Recherche du 
Centre National de la Recherche Scientifique et de 
l' Ecole Normale Sup{\'e}rieure. } }
\centerline{ 24 rue Lhomond, 75231 Paris CEDEX 05, France. }
\centerline { e-mail: $\ $ sourlas@lpt.ens.fr }

{\vskip 24pt}

\noindent{\eightbf Abstract:} {\eightrm I will show that there is a deep relation between error-correction
codes and certain mathematical models of spin glasses.
 In particular minimum error probability decoding is equivalent to
finding the ground state  of the corresponding spin system.
The most probable value of a symbol is related to the magnetization
at a different temperature. Convolutional codes correspond to one-dimensional
spin systems and Viterbi's decoding algorithm to the transfer matrix
algorithm of Statistical Mechanics.
 
I will also show how the recently discovered (or rediscovered) 
capacity approaching codes (turbo codes and 
 low density parity check codes) can be analysed
using statistical mechanics. It is possible to show, using statistical
mechanics, that these  codes allow error-free communication for signal to
noise ratio above a certain threshold. This threshold depends on the
particular code, and can be computed analytically in many cases.  
}

{\vskip 12pt}

\noindent LPTENS 01/ 31

%\noindent{\eightbf Keywords:} {\eightrm List of keywords.}

{\vskip 15pt}

 It has been known\ref{1-4} that error-correcting codes are 
mathematically equivalent to some theoretical spin-glass models. 
As it is explained in Forney's paper in this volume, 
 there have been recently 
very interesting new developments in error-correcting codes. It 
is now possible to approach practically very close to Shannon's channel 
capacity. First came the discovery of turbo codes by Berrou and Glavieux\ref{5} 
and later the rediscovery of low density parity check codes\ref{6}, 
first discovered by Gallager\ref{7,8}, in his thesis, in 1962. 
Both turbo codes and low density parity check (LPDC) codes are based 
on random structures. It turns out, as I will explain later, that 
it is possible to use their equivalence with spin glasses, 
to analyse them using the methods of statistical mechanics.

Let me start by fixing the notations.  
 Each information message consists of a sequence of $K$ bits 
$\vec u = \{ u_{1}, \cdots , u_{K} \}, u_{i} = 0$ or $ 1 $. 
The binary vector $\vec u $ is called the source-word. 
Encoding introduces redundancy into the message. 
One maps $\vec u \to \vec x $ 
by encoding. $\vec u \to \vec x $ has to be a one to one map for the 
code to be meaningful. The binary vector $\vec x $ has $N > K $ components. 
It is called a code-word. 
The ratio $ R = K / N $ which specifies the redundancy of the code, 
is called the rate of the code. 
One particularly important family of codes are the so-called 
linear codes. Linear codes  are defined by 
$$\vec x = G \vec u $$
 $G$ is a binary (i.e; its elements are zero or one) 
$(N \times K)$ matrix and the multiplication is modulo two. 
$G$ is called the generating matrix of the code. 
Obviously by construction all the components $x_{i}$ of a code-word $x$ are 
not independent. 
Of all the $2^N$ binary vectors only $2^{K} = 2^{NR} $, those 
corresponding to a  vector $ \vec u $, are code-words. 
Codewords satisfy the linear constraints (called parity check constraints) 
$ H \vec x = 0 $ (modulo two), 
where $H$ is a $(K \times N)$ 
binary matrix, called the parity check matrix. 
The connection with spin variables is straightforward. 
$ u_{i} \to \sigma_{i} = (-1)^{u_{i}} $, $ x_i \to J_{i} = (-1)^{x_{i}} $. 
It follows that $ u_i + u_j =  \sigma_{i}  \sigma_{j} $ and 
$$ J_{i} = (-1)^{\sum_j G_{ij} u_{j}} = C^{i}_{k_{1} \cdots k_{i}} 
\sigma_{k_{1}} \cdots \sigma_{k_{i}} \eqno(1) $$ 
The previous equation defines the ``connectivity matrix '' 
$C$ in terms of the generating matrix of the code $G$. 
Similarly one can write the parity check constraints in the form: 
$$ (-1)^{ \sum_{j} H_{lj} x_{j} } = 1 \ \to M^{l}_{k_{1} \cdots k_{l}} 
J_{k_{1}} \cdots J_{k_{l}} =1 \eqno(2) $$
This defines the ``parity constraint matrix '' 
$M$ in terms of the parity check matrix $H$ of the code.

Codewords are sent through a noisy transmission 
channel and they get corrupted because of the channel noise. 
 If a $ J_i  = \pm 1 $ is sent, the output will be different, 
in general a real number $ J^{out}_{i}$. 
 Let us call 
$ Q ( \vec J^{out} | \vec J ) d \vec J^{out} $ the probability for the 
transmission channel's output to be between $\vec J^{out} $ and $ \vec J + d \vec J^{out}$,
 when the input was $\vec J $. 
The channel ``transition matrix'' $ Q ( \vec J^{out} | \vec J ) $  
is supposed to be known. We will assume that 
the noise is independent for any pair of bits (``memoryless channel"), i.e. 
$$ Q ( \vec J^{out} | \vec J ) = \prod_{i} q ( J^{out}_i | J_i ) \eqno(3) $$
 Communication  is a statistical inference problem. 
Knowing the noise  probability i.e. $q ( J^{out}_i | J_i ) $, 
the code (i.e. in the present case of linear codes 
knowing the generating matrix $G$  or the parity check matrix $H$) and 
the channel output $\vec J^{out} $, one has 
 to infer the message that was sent.
The quality of inference depends on the choice of the code. 

We will now show that there exists a close mathematical relationship 
between error-correcting codes and theoretical models of disordered 
systems. To every possible information message 
(source word) $ \vec \tau $ we can assign a probability 
$P^{source}( \vec \tau | \vec J^{out} ) $, conditional on the channel 
output $\vec J^{out} $. Or, equivalently, to any 
code-word $\vec J $ we can assign a probability 
$ P^{code}( \vec J | \vec J^{out} ) $. 

Because of Bayes theorem, the probability for any code-word 
symbol (``letter'') $J_i = \pm 1 $, $p(J_i | J^{out}_i) $,  
conditional on the channel output $J^{out}_i $, is given by 
$$ \ln p(J_i | J^{out}_i) \  = \ c1 \ + \ \ln q( J^{out}_i | J_i )  
\  = \ c2 \ + \ h_i J_i \eqno(4) $$ 
where $c1$ and $c2$ are constants (non depending on $ J_i $) and 
$$ h_i \  = \ { 1 \over 2 } \ln { q(J^{out}_i | + 1 ) \over q( J^{out}_i | -1 ) }  \eqno(5) $$
It follows that 
$$ P^{code}( \vec J | \vec J^{out} ) = c  \prod_{l} \delta (
M^{l}_{k_{1} \cdots k_{l}} J_{k_{1}} \cdots J_{k_{l}} ; 1 ) \exp { 
(\sum_i  h_i J_i )} \eqno(6) $$ 
where $c$ is a normalising constant. 
The Kronecker $\delta$'s enforce the constraint that $ \vec J $ 
obeys the parity check equations (Equ. (2) ), i.e. that it is a code-word. 
The $\delta$'s can be replaced by a soft constraint, 
$$ P^{code}( \vec J | \vec J^{out} ) = const \exp 
{\ [ u \  \ \sum_l M^{l}_{k_{1} \cdots k_{l}} J_{k_{1}} \cdots J_{k_{l}}  
\ + \ \sum_i  h_i J_i  \ ] } \eqno(7) $$
where $ u \to \infty $. 
We now define the corresponding spin Hamiltonian by: 
$$ -H^{code} (\vec J) = \ln P^{code}( \vec J | \vec J^{out} ) = 
u \sum_l M^{l}_{k_{1} \cdots k_{l}} J_{k_{1}} \cdots J_{k_{l}} + 
\sum_i  h_i J_i  \eqno(8) $$
This is a spin system with multispin 
interactions and an infinite 
ferromagnetic coupling and a random external magnetic field. 

Alternatively, one may proceed by solving the parity check 
constraints 
$ J_{i} = C^{i}_{k_{1} \cdots k_{i}} 
\sigma_{k_{1}} \cdots \sigma_{k_{i}} $ (i.e. by expressing the 
code-words in terms of the source-words). 
$$ P^{source}( \vec \sigma | \vec J^{out} ) = const. \  \exp { 
(\sum_i  h_i C^{i}_{k_{1} \cdots k_{i}} 
\sigma_{k_{1}} \cdots \sigma_{k_{i}} )} \eqno(9) $$ 
where the $h_i$'s are given as before. The logarithm of 
$ P^{source}( \vec \sigma | \vec J^{out} ) $, 
$$ H^{source}(  \vec \sigma ) = - \ln P^{source}( \vec \sigma | \vec J^{out} ) 
= - \sum_i  h_i C^{i}_{k_{1} \cdots k_{i}} 
\sigma_{k_{1}} \cdots \sigma_{k_{i}}  \eqno(10) $$
 has obviously the form of a 
spin glass Hamiltonian. 

We have given two different statistical mechanics formulations of error 
correcting codes. One in terms of 
the souceword probability $P^{source}$ and the other in terms of 
 the code-word probability $P^{code}$.   
 Because of the one to one correspondence 
between code-words and source-words, the two formulations  
are equivalent. In practice however it may make a difference. 
It may be more convenient to work with $P^{source}$ or $P^{code}$, 
depending on the case. For the case of turbo codes (see later) 
it will be more convenient to define another probability, 
the ``register word'' probability.

It follows that the most probable sequence (``word MAP decoding'') 
is given by the 
ground state of this Hamiltonian ($ H^{code} $   or $ H^{source}  $, 
depending on the case). Instead of considering the most 
probable instance, 
one may only be interested in 
the most probable value  $\tau_{i}^{p} $  of the i'th 
``bit" $\tau_i $\ref{9,10,11} (``symbol MAP decoding'').
 Because $\tau_i = \pm 1 $, the probability $p_i$ for $\tau_{i} =1 $ 
is simply related to $ m_i $, the average  of $\tau_{i} $,
$p_i = (1+m_i)/2 $. 

$$ m_i  \ = \ { 1 \over Z } \ \sum_{ \{ \tau_1 \cdots \tau_N \} } 
 \tau_{i} \ \exp -  H (\vec \tau )  \quad 
 Z = \sum_{ \{ \tau_1 \cdots \tau_N \} } 
  \ \exp -  H (\vec \tau )\quad \tau_{i}^{p}  = 
{\rm sign} \ ( m_i )   \eqno(11) $$
In the previous equation $m_i$ is obviously the thermal average 
at temperature $T=1$.
It is amusing to notice that
 $T=1$ corresponds to Nishimori's temperature\ref{12}. 

When all messages are equally probable and the transmission channel is 
 memoryless and symmetric, i.e. when $q ( J^{out}_i | J_i ) = 
q ( -J^{out}_i | -J_i ) $, the error 
probability is the same for all input sequences. It is enough to compute 
it in the case where all input bits are equal to one, 
i.e. when the transmitted code-word is the all zero's code-word. 
In this case, the error probability per bit $P_{e} $ is 
$ P_{e} \ = \ { 1 - m^{(d)} \over 2 }$, where 
$ m^{(d)} \ = \ { 1 \over N} \sum_{i=1}^{N} \tau_{i}^{(d)} $ and 
$ \tau_{i}^{(d)} $ is the symbol sequence produced by the decoding procedure.

This means that it is possible to compute the bit error probability, 
if one is able to compute the magnetization in the corresponding 
spin system. 

Let me give a simple example of an $ R = 1/2 $ ``convolutional'' code. 
From the $ N $ source symbols (letters) $u_i$'s we construct the 
$2N$ code-word  letters 
 $x_{k}^{1} $, 
$x_{k}^{2} $, $k=1, \cdots , N $.
$$ x_i^{1} = u_{i}+u_{i-1}+u_{i-2} \ , \ \ 
x_i^{2} = u_{i}+u_{i-2} \eqno(12) $$
It follows that 
$$ J_{k}^{1} = \sigma_{k}  \sigma_{k-1}  \sigma_{k-2} \ , \ \ 
J_{k}^{2} = \sigma_{k}  \sigma_{k-2} \eqno(13) $$ 
$$ C^{(1,k)}_{i_{k_{1}} i_{k_{2}} i_{k_{3}}} = \delta_{k,i_{k_{1}}}  
 \delta_{k,i_{k_{2}}+1 }  \delta_{k,i_{k_{3}}+2} \ , \ \ 
 C^{(2,k)}_{i_{k_{1}}  i_{k_{3}}} = \delta_{k,i_{k_{1}}} 
 \delta_{k,i_{k_{3}}+2}  \eqno(14) $$
The corresponding spin Hamiltonian is 
$$ -H = { 1 \over w^2 } \sum_k J_{k}^{1,out} \tau_{k}  \tau_{k-1}  
\tau_{k-2} + J_{k}^{2,out} \tau_{k}    \tau_{k-2} \eqno(15) $$
Here I assumed a Gaussian noise. In that case, Equ. (5) 
reduces to $h_k = J_{k}^{out} / w^2 $, 
where $ w^2 $ is the variance of the noise. 
This is a one dimensional spin glass Hamiltonian. In fact it is easy to 
see that convolutional codes correspond to one dimensional 
spin systems. Their ground state 
can be found using the $T=0$ transfer matrix algorithm. This corresponds to 
the Viterbi algorithm in coding theory. 
For symbol MAP (maximum a posteriori probability) decoding, one 
can use the $T=1$ transfer matrix algorithm. This in turn is the 
BCJR algorithm in coding theory\ref{13}.

As it is explained in Forney's paper in this volume, the newly discovered 
(or rediscovered) capacity approaching codes are based on random 
constructions. Using the equivalence explained above it has been 
 possible to analyse them using the methods  
of statistical mechanics. 

Gallager's low density parity check ($ k , l $) codes are defined 
by choosing at random a sparse parity check matrix $H$ as follows. 
$H$ has $N$ columns (we consider the case of code-words of length $N$). 
 Each column of $H$ has $k$ elements equal to one and all other 
elements equal to zero. Each row has $l$ non zero elements. It 
follows that $H$ has $N k/l$ rows and that 
the rate of the code is $R=1-k/l$. It follows from equation (8) 
that Gallager's $ k , l $ codes correspond to diluted 
spin models with $l$-spin infinite strength ferromagnetic interactions 
in an external random field. It turns out 
that the belief propagation algorithm, 
used to decode LPDC codes, amounts to an iterative solution 
of the Thouless Anderson Palmer\ref{14} (TAP) equations for spin glasses. 
 A detailed analysis of these codes 
is presented in Urbanke's paper in this volume. 
Low density parity check codes have been analysed using 
Statistical Mechanics methods by Kabashima Kanter and Saad\ref{15,16} in the 
replica symmetric approximation. 
More recently Montanari\ref{17} was able to establish the entire phase 
diagram of LDPC codes. For $k, \ l \to \infty$ with $k/l$ fixed, he showed 
that $ k , l $ codes correspond to a random energy model 
which can be solved without replicas. There is a phase transition in this 
model, which occurs at a critical value 
of the noise $n_c$. $n_c$ separates a zero error phase from a high 
error phase. $n_c$ in this case equals the value provided by 
Shannon's channel capacity. For finite $k$ and $l$ he found an 
exact one step replica symmetry breaking solution. The 
location of the phase transition determines $n_c$. In this way 
he computed also for finite values of $k$ and $l$ the critical value of the 
noise below which error free communication is possible. 
A different value of $n_c$, $n_c^{bp}$ had already being computed 
by Richardson and Urbanke\ref{18} (see  Urbanke's paper in this volume). 
Richardson and Urbanke compute $n_c^{bp}$ by analysing the behaviour 
of the decoding algorithm, belief propagation in this case. 
Statistical mechanics provides a threshold $n_c$ which 
in principle is different 
from $n_c^{bp}$. $n_c$ is reached  by the optimum (but unknown) decoder. 

Turbo Codes also have been analysed using statistical mechanics\ref{19,20}.
Turbo Codes are based on recursive convolutional codes. An example of 
non recursive convolutional code was given in Equ. (12). The 
corresponding recursive code is given, most conveniently, in terms 
of the auxiliary bits $b_i$, defined below. The $b_i$'s are 
stored in the encoder's memory registers, that's why 
I call $ \vec b $ the the ``register word''.  
$$x_i^{1} = u_{i}, \  x_i^{2}  = b_{i}+b_{i-2},   \ b_{i} = 
u_{i}+b_{i-1}+b_{i-2}  \eqno(16) $$
It follows that the source letters $u_i$ are given in terms of the 
 auxiliary ``register letters'' $b_i$ 
$$ u_i = b_{i}+b_{i-1}+b_{i-2} \eqno(17) $$
All additions are modulo two. 

To construct a turbo code, one 
artificially considers a second source word $ \vec v $, by 
performing a permutation, chosen at random, on the 
original code-word $ \vec u $. So one considers 
$v_i = u_{P(i)} $ where $j = P(i) $ is a (random) permutation 
of the $K$ indices $i$ and a second ``register word'' $c_i$, 
$ c_i = v_{i}+c_{i-1}+c_{i-2}$. Obviously 
$$ v_i = c_i +c_{i-1}+c_{i-2} = u_j = b_{j}+b_{j-1}+b_{j-2},  \ \ 
j = P(i)  \eqno(18) $$
Equ. (18) can be viewed as a constraint on the two register words  
$\vec b $ and $\vec c $. Finally in the present example, a rate 
$R=1/3$ turbo code, one transmits the $3K$ letter code-word 
$x_i^{1} =  u_{i}$, $ x_i^{2} = b_{i}+b_{i-2} $,    
 $ x_i^{3} = c_{i}+c_{i-2} $, $i=1, \cdots, K$. 
Let's call, as before,  
$$J_i^{\alpha}=(-1)^{x_{i}^{\alpha}} , \ \ \alpha=1,2,3 $$ 
the channel inputs and $J_i^{out,\alpha} $ the channel outputs. 
In the previous, for reasons of convenience, 
 we formulated convolutional codes using  
 the source-word probability $P^{source}$ and LDPC codes using the 
code-word probability  $P^{code}$. 
The statistical mechanics of turbo codes is most conveniently formulated 
in terms of the ``register words'' probability 
$P^{reg}( \vec \sigma , \vec \tau | \vec J^{out} ) $ conditional on 
the channel outputs $ \vec J^{out} $, where $ \tau_i = (-1)^{b_i} $ 
and $ \sigma_i = (-1)^{c_i} $. The logarithm of this probability 
provides the spin Hamiltonian 
$$- H = {1 \over w^2 } \sum_k  J_{k}^{out,1} \tau_{k}  \tau_{k-1}  
\tau_{k-2} + J_{k}^{out,2} \tau_{k}    \tau_{k-2} 
+ J_{k}^{out,3} \sigma_{k}    \sigma_{k-2}    \eqno(19) $$
Because of Equ. (18), the two spin chains $\vec \tau $ and 
$\vec \sigma $ obey the constraints 
$$ \sigma_i \sigma_{i-1} \sigma_{i-2} = \tau_{j} \tau_{j-1} 
\tau_{j-2} ,  \ \ j = P(i)  \eqno(20) $$
(As previously, we have considered the case of a Gaussian noise 
of variance $w^2$.) 
This is an unusual spin Hamiltonian. Two short range one dimensional 
chains are coupled through the infinite range, non local 
 constraint, Equ. (20). This constraint is non local because neighboring 
$i$'s are not mapped to neighboring $j$'s under the random permutation. 
It turns out that this Hamiltonian can be solved by the replica method. 
One finds a phase transition at a critical value of the noise 
$n_{crit}$. For noises less than $n_{crit}$, the magnetization 
equals one, i.e. it is possible to communicate error free. In this respect, 
turbo codes are similar to Gallager's LDPC codes. The statistical 
mechanical models however, are completely different. 
Let me also mention that, under some reasonable assumptions, 
the iterative decoding algorithm for turbo codes 
(turbodecoding algorithm), which I am not 
explaining here, can be viewed\ref{20} as a time discretisation of 
the Kolmogorov, Petrovsky and Piscounov equation\ref{21}. This KPP 
equation has traveling wave solutions. The velocity of the traveling 
wave, which is computable analytically, corresponds to the 
convergence rate of the turbodecoding algorithm. The 
agreement with numerical simulations is excellent.

So the equivalence between linear codes 
and theoretical models of spin glasses is 
quite general and we have established the following 
dictionary of correspondence.
$$\eqalign { Error-correcting \ code \ &\iff \ Spin \ Hamiltonian \cr
Signal \ to \ noise \ &\iff \ J_{0}^2/\Delta J^2 \cr
Maximum \  likelihood \ Decoding\ &\iff \ Find \ a \ ground \ state \cr
Error \ probability \ per \ bit \ &\iff \ Ground \ state \ magnetization \cr
Sequence \  of \  most \  probable  \ symbols &\iff \ magnetization \  
 at  \  temperature \   T=1 \cr
Convolutional  \  Codes &\iff \ One \   dimentional  \  spin-glasses  \cr
Viterbi  \  decoding  &\iff \ T=0 \ Transfer \   matrix  \  algorithm  \cr 
BCJR  \  decoding  &\iff \ T=1 \  Transfer \   matrix  \  algorithm  \cr 
Gallager  \  LDPC  \  codes  &\iff Diluted \ p-spin \ ferromagnets 
\ in \ a \ random \ field    \cr
Turbo  \  Codes    &\iff Coupled  \ spin \ chains PC    \cr
Zero \ error \ threshold  &\iff  Phase \ transition \ point \cr 
Belief \ propagation \ algorithm &\iff  Iterative \ solution \ of \ TAP \ 
equations  } $$

I would like to conclude by pointing out some open questions. 

What is the order of the phase transition? This question is 
particularly relevant for turbo codes and has important implications 
for decoding.

What are the finite size effects? This question is particularly 
relevant near the zero error noise threshold (i.e. near the phase 
transition). The answer will depend on the order of the transition. 

How does  the decoding complexity behave as one approaches the 
zero error noise threshold? Is there a critical slowing down? 
As it was said before, the decoding algorithms both for 
LDPC codes and turbo codes are heuristic and there are not known results 
as one approaches the phase transition. 

Is there a glassy phase in decoding? In other terms, do the  heuristic 
decoding algorithms reach the threshold of optimum decoding, 
computed by statistical mechanics, or is there a (lower) noise 
``dynamical'' threshold where decoding stops reaching optimal 
performance?

I hope that at least some of the above questions will be 
answered in the near future.

\vfill\eject

\vglue .7 true cm
\centerline{\bf References }
\vglue .6 true cm

\item {1)} Sourlas, N., {\it Nature} 339, 693 (1989)

\item {2)} Sourlas, N., in {\it Statistical Mechanics of Neural Networks}, 
Lecture Notes in Physics 368, ed. L. Garrido, Springer Verlag (1990)

\item {3)} Sourlas, N., Ecole Normale Sup\'erieure preprint (April 1993)

\item {4)} Sourlas, N., in {\it From Statistical Physics to 
 Statistical Inference and Back,}
 ed. P. Grassberger and J.-P. Nadal, Kluwer Academic (1994) p. 195.

\item {5)} C. Berrou, A. Glavieux, and
P.Thitimajshima. Proc.1993 Int.Conf.Comm. 1064-1070

\item {6)} MacKay, D. J. C. Neal, R. M. {\it Elect. Lett.}  
33, 457 (1997).

\item {7)} Gallager, R. G.  {\it IRE Trans. Inform. Theory },
 IT-8, 21 (1962).

\item {8)} Gallager, R. G.  {\it Low-Density Parity-Check Codes },
 MIT Press, Cambridge MA (1963).

\item {9)} Ruj\'an, P., {\it Phys. Rev. Lett.} 70, 2968 (1993)

\item {10)} Nishimori, H., {\it J. Phys. Soc. Jpn. }, 62, 2973 (1993)

\item {11)} Sourlas, N., {\it Europhys. Lett.} 25, 169 (1994)

\item {12)} Nishimori, H., {\it Progr. Theor. Phys.} 66, 1169 (1981)

\item {13)} L. Bahl, J. Cocke, F. Jelinek, and J. Raviv. IEEE
Trans.Inf.Theory {\bf IT-20}(1974) 248-287

\item {14)} Thouless, D. J. Anderson, P. W. Palmer, R. G. 
{\it Phil. Mag. } 35, 593 (1977)

\item {15)} Kanter, I. and Saad, D. {\it Phys. Rev. Lett.} 83, 2660 (1999)

\item {16)} Kabashima, Y. Murayama T. and Saad, D. 
{\it Phys. Rev. Lett.} 84, 1355 (2000)

\item {17)} Montanari, A. cond-mat/0104079

\item {18)} Richardson, T. J. Urbanke, R. L. {\it IEEE Trans. 
Inform. Theory } 47, 638 (2001).

\item {19)} Montanari, A. Sourlas, N. {\it Eur. Phys. J. } B 18, 
107 (2000) 

\item {20)} Montanari, A. {\it Eur. Phys. J. } B 18, 121 (2000)

\item {21)} Kolmogorov, A. Petrovsky, I and Piscounov, N. 
{\it Moscou Univ. Math. Bull. } 1, 1 (1937). 
 
\end